\documentclass[fleqn,12pt,twoside]{article}
\usepackage{graphicx}
\title{Behind the Success of the Quark Model}

\author{
T.~T.~Takahashi
\thanks{RCNP, Osaka University, Mihogaoka 10-1, Ibaraki, Osaka 567-0047, Japan},
H. Suganuma\thanks{Faculty of Science, Tokyo Institute of Technology, Tokyo 152-8551, Japan},
H. Ichie\thanks{Humboldt Univ. zu Berlin, Inst. f\"ur Phys., Invalidenstrasse, D-10115 Berlin, Germany},
H. Matsufuru\thanks{Yukawa Institute for Theoretical Physics, Kyoto University, Kyoto 606-8502, Japan}
and Y. Nemoto\thanks{RIKEN-BNL Research Center, Brookhaven National Laboratory, Upton 11973, USA}
}
       
\begin{document}

\maketitle

\begin{abstract}
The ground-state three-quark (3Q) potential $V_{\rm 3Q}^{\rm g.s.}$
and the excited-state 3Q potential $V_{\rm 3Q}^{\rm e.s.}$
are studied using SU(3) lattice QCD at the quenched level.
For more than 300 patterns of the 3Q systems, 
the ground-state potential $V_{\rm 3Q}^{\rm g.s.}$ is investigated in detail 
in lattice QCD with $12^3\times 24$ at $\beta=5.7$ and
with $16^3\times 32$ at $\beta=5.8, 6.0$.
As a result, the ground-state potential $V_{\rm 3Q}^{\rm g.s.}$
is found to be well described with Y-ansatz within the 1\%-level deviation.
From the comparison with the Q-$\rm\bar Q$ potential, we find 
the universality of the string tension as 
$\sigma_{\rm 3Q}\simeq\sigma_{\rm Q\bar Q}$ 
and the one-gluon-exchange result as $A_{\rm 3Q}\simeq\frac12 A_{\rm Q\bar Q}$.
The excited-state potential $V_{\rm 3Q}^{\rm e.s.}$
is also studied in lattice QCD with $16^3\times 32$ at $\beta=5.8$ 
for 24 patterns of the 3Q systems.
The energy gap between $V_{\rm 3Q}^{\rm g.s.}$ and $V_{\rm 3Q}^{\rm e.s.}$, 
which physically means the gluonic excitation energy, is found to be about 1GeV 
in the typical hadronic scale, 
which is relatively large compared with the excitation energy of the quark origin. 
This large gluonic excitation energy justifies the great success of the simple quark model.
\end{abstract}

\section{Introduction}
\label{I}
The hadron dynamics is ruled by quantum chromodynamics (QCD).
In a spirit of the elementary particle physics, it is desired 
to comprehend the hadron system directly from QCD.
However, due to its strong-coupling nature, it is still difficult
to extract the hadron dynamics from QCD in an analytic manner.
Instead, the lattice QCD calculation has been adopted as a promising method
for the nonperturbative analysis of QCD.

In the first half of this paper, we study the three-quark (3Q) potential in lattice QCD. 
The inter-quark potential is one of the most fundamental and important
quantities in QCD, and is directly responsible for the hadron properties.
In contrast with lots of lattice studies for the quark-antiquark (Q-$\rm\bar Q$) potential,
there had been no reliable lattice-QCD result on the 3Q potential, although 
the 3Q potential is responsible for the baryon properties  
and of great importance also for the quark-confinement mechanism in baryons.
On the ground-state 3Q potential $V_{\rm 3Q}^{\rm g.s.}$,
the detailed study has been recently performed, and Y-ansatz is now almost conclusive~\cite{TMNS01,TSNM02}.

In the latter half of this paper, we consider the connection between QCD and the quark model
in terms of the excited-state inter-quark potential. 
The low-lying hadron properties, especially for baryons,
can be successfully reproduced in 
the framework of the simple non-relativistic quark model~\cite{RGG75}, which 
has only quark degrees of freedom and has no gluonic modes.
The non-relativistic treatment can be justified
by spontaneous chiral-symmetry breaking, which gives rise to
a large constituent quark mass of about 300 MeV.
On the other hand, we have no reason
which supports the absence of the gluonic excitation modes 
in the low-lying hadron spectra.
To give a solution to this mystery, we investigate the gluonic excitation modes.
In spite of several lattice studies on the gluonic excitation mode
in the Q-$\rm\bar Q$ system, the gluonic excited-state potential
in the 3Q system has not been investigated in lattice QCD. 
We show the first result on  
the 1st excited-state potential $V_{\rm 3Q}^{\rm e.s.}$
in the spatially-fixed 3Q system in SU(3) lattice QCD~\cite{TS02}.

\section{The ground-state 3Q potential -- Lattice QCD evidences of Y-ansatz}
\label{II}

The Q-$\rm\bar Q$ potential is described by a simple form as 
$V_{\rm Q\bar Q}^{\rm g.s.}(r)=-\frac{A_{\rm Q\bar Q}}{r}
+\sigma_{\rm Q\bar Q}r+C_{\rm Q\bar Q}$.
In a theoretical conjecture, due to the confinement effect at the large distance, 
the color-electric flux among the quarks is expected to be squeezed as the color-flux-tube, and 
the quarks are linked with the one-dimensional flux-tube, 
which leads to the linear potential.
Reflecting the SU(3) gauge symmetry in QCD, 
the color-flux-tube in the 3Q system has a junction which connects
three different colors in a color-singlet manner.
In the ground state of the 3Q system, 
the flux-tube energy proportional to its length is minimized in the presence of the junction, 
and therefore the Y-type flux-tube is expected to be formed among the three quarks~\cite{TMNS01,TSNM02,I02}.
Thus, similarly in the Q-$\rm\bar Q$ potential, 
the 3Q potential\cite{TMNS01,TSNM02} is conjectured 
to be expressed by a sum of a constant, 
the two-body Coulomb term from the perturbative OGE process 
at the short distance and 
the three-body linear confinement term at the long distance: 
\begin{eqnarray}
V_{\rm 3Q}=-A_{\rm 3Q}\sum_{i<j}\frac1{|{\bf r}_i-{\bf r}_j|}
+\sigma_{\rm 3Q} L_{\rm min}+C_{\rm 3Q} \qquad \hbox{(Y-ansatz)}
\label{3Qp}
\end{eqnarray}
with the minimal value $L_{\rm min}$ of the total length of 
color-flux-tubes linking three quarks. 

\begin{figure}
\begin{center}
\includegraphics[scale=1.2]{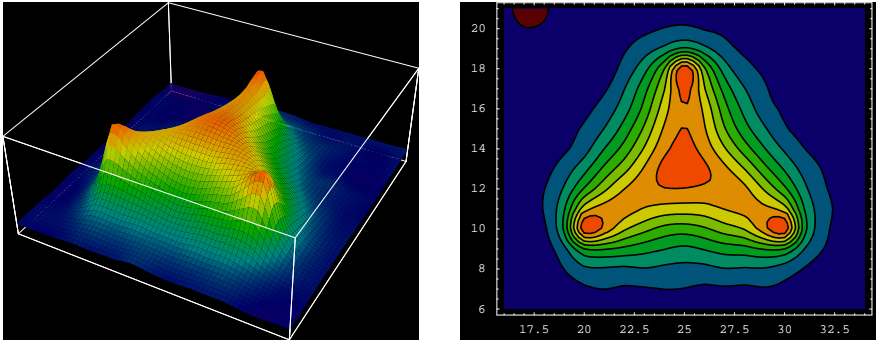}
\vspace{-0.5cm}
\caption{\label{flux}
The lattice QCD result for the flux-tube profile 
in the spatially-fixed 3Q system in the maximally-abelian projected QCD~\cite{I02}.
The distance between the junction and each quark is about 0.5 fm.
}
\end{center}
\vspace{-0.75cm}
\end{figure}

We extract the static 3Q potential $V_{\rm 3Q}^{\rm g.s.}$
from the 3Q Wilson loop \cite{TMNS01,TSNM02}
using SU(3) lattice QCD calculations
in the model-independent way.
For the accurate measurement of $V_{\rm 3Q}^{\rm g.s.}$, 
we adopt the gauge-covariant smearing method, which enhances the ground-state overlap.
For more than 300 patterns of the 3Q systems, 
we investigate $V_{\rm 3Q}^{\rm g.s.}$ in lattice QCD 
with $12^3\times 24$ lattice at $\beta=5.7$ and
with $16^3\times 32$ lattices at $\beta=5.8, 6.0$.

As a result, the 3Q potential can be accurately described
by the Y-ansatz form in Eq.(\ref{3Qp}) within the 1\%-level deviation.
From the comparison with the Q-$\rm\bar Q$ potential,
we find the universality of the string tension as
$\sigma_{\rm 3Q}\simeq\sigma_{\rm Q\bar Q}$ and
the one-gluon-exchange result as $A_{\rm 3Q}\simeq\frac12 A_{\rm Q\bar Q}$.

This result is also supported by the recent lattice study 
on the flux-tube profile in the 3Q system~\cite{I02}.
The Y-type flux-tube is observed in the 3Q system as shown in Fig.1.
(In contrast, the theoretical basis of the $\Delta$-ansatz in Ref.\cite{C96} 
insists on the absence of any flux-tube formation in the 3Q system: 
neither Y nor $\Delta$ flux-tube is to be observed in their framework.)
Thus, the possibility of $\Delta$-ansatz~\cite{C96} is almost denied, 
and Y-ansatz seems established for the 3Q potential.

\section{The excited-state 3Q potential and the success of the simple quark model} 
\label{TES3QP}

Color confinement mechanism in QCD leads to a squeezed color-flux-tube, 
and then the gluonic excitation is theoretically expected to appear as the vibrational 
mode of the color-flux-tube~\cite{TS02}. 
Experimentally, this type of the gluonic excitation is 
closely related to the hybrid mesons and the hybrid baryons, 
which consist of $q\bar qG$ and $qqqG$, respectively, in the valence picture.
In particular, the several states with the exotic quantum number such as $J^{PC}=0^{--},0^{+-},1^{-+},2^{+-},...$ 
cannot be constructed with the simple quark picture.

As for the excited-state Q-$\rm\bar Q$ potential, recent lattice studies indicate the form as 
$V_{\rm Q\bar Q}^{\rm e.s.}(r)=\sqrt{b_0+b_1r+b_2r^2}+c_0$ \cite{JKM98}.
However, the vibrational modes of the Y-type flux-tube system are considered to be 
much more complicated and 
chaotic than those of the simple Q-$\bar{\rm Q}$ flux-tube, 
because of the possible interference 
among the vibrations of the three flux-tubes 
connected at the physical junction.

\begin{figure}
\begin{center}
\includegraphics[scale=0.50]{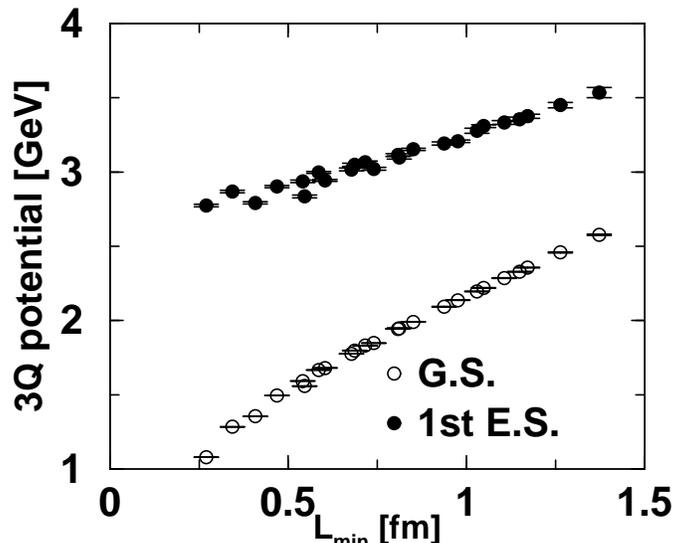}
\vspace{-0.75cm}
\caption{\label{fluxandexcited}
The lattice QCD results of 
the ground-state 3Q potential $V^{\rm g.s.}_{\rm 3Q}$ (open circles) and the 1st 
excited-state 3Q potential $V^{\rm e.s.}_{\rm 3Q}$ (filled circles) as the function of $L_{\rm min}$.
The gluonic excitation energy is found to be more than 1GeV in the hadronic scale.
}
\end{center}
\vspace{-0.75cm}
\end{figure}

We investigate the 1st excited-state potential in the spatially-fixed
3Q system using lattice QCD with $16^3\times 32$ lattice at
$\beta$=5.8 for 24 patterns of the 3Q systems.
In Fig.2, we plot the ground-state potential $V^{\rm g.s.}_{\rm 3Q}$ and 
the 1st excited-state potential $V^{\rm e.s.}_{\rm 3Q}$ as the function of $L_{\rm min}$ 
in the physical unit.

As a remarkable fact, the lowest gluonic excitation energy $\Delta E=V^{\rm e.s.}_{\rm 3Q}-V^{\rm g.s.}_{\rm 3Q}$ 
is found to be about 1GeV 
in the hadronic scale as $L_{\rm min} \simeq 0.5-1.5{\rm fm}$.
This is rather large in comparison with the low-lying excitation energy
of the quark origin.
(Also for the Q-${\rm \bar Q}$ system, a large gluonic excitation energy is reported in recent lattice studies~\cite{JKM98}.)
Such a gluonic excitation contribution would be significant in 
the highly-excited baryons with the excitation energy above 1GeV, and  
the lowest hybrid baryon~\cite{CP02}, which is described as $qqqG$ in the valence picture, 
is expected to have a large mass of about 2GeV.

The large gluonic excitation energy corresponding to the flux-tube vibrational energy is considered to 
originate from  the ``tight'' flux-tube with a large string tension, 
as a result of the strong confinement effect. 
Due to the large gluonic excitation energy $\Delta E \simeq$ 1GeV, 
the gluonic excitation modes are invisible in 
the low-lying excitations of hadrons, which can be considered as
a reason of the success of the simple quark model without gluonic modes~\cite{TS02}.
(See Fig.3.)

\begin{figure}
\begin{center}
\includegraphics[scale=0.75]{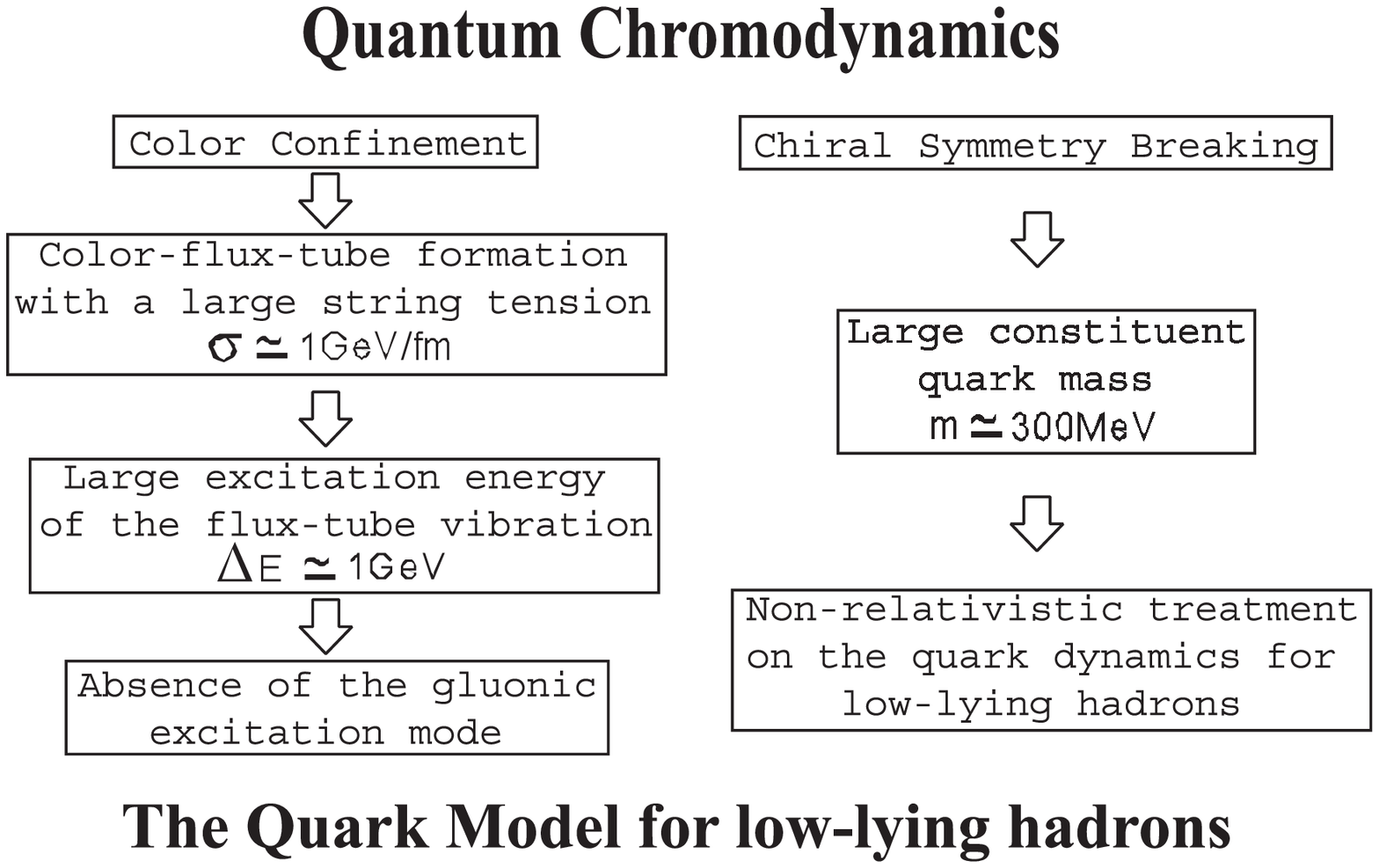}
\vspace{-0.75cm}
\caption{\label{behind}
Connection from QCD to the success of the quark model for low-lying hadrons.
The large gluonic excitation energy $\Delta E \simeq$ 1GeV leads to 
the absence of the gluonic mode in the low-lying hadrons 
and brings about the great success of the quark model.
}
\end{center}
\vspace{-0.75cm}
\end{figure}

\end{document}